# Reality: Physics; "Physical Universe Language" and Mathematics; Physics Formalisms in Human Brain "Machine Language"

J. N. Schad, PhD

Retired LBNL (UCB) Scientist

schadn5@berkeley.edu

Abstract

The nature of the existence, revealed through Human's cognitive system, has been evolving since the development of the languages. Part of such revelations, appearing early in the history of the civilization, were the geometrical forms and the numbers, whose beauty and orders, wondrous and mysterious, conveyed a sense of unreality, beyond the physical reality normally discerned; to Plato they were indications of the "other reality," of which only those few glimpses had occurred. And full access to it, for the earthlings, required breaking out of the shackles of their (mental) captivity; beautifully portrayed in the Master's cave allegory. The Present era's sense of wonder of the effectiveness of the mathematics in the formulations of the discoveries of the fundamental and natural laws of existence, on the one hand, and on the other, the advances in computation, and especially in AI, have revived the platonic idea, in one form or other, with some ardent adherents among the very minds who contribute to furthering of our understanding of our physical universe. The work presented here is an attempt in proving that a "physical world," precludes the prospects for the "mathematics" to be its "other reality." This end is achieved by bringing to light the processes involved in the perceptions and discernment of the world by humans, and for that matter, for any creature with a physical nervous system. The fundamentals involved make it clear that the events of the "physical universe" played out in our "physical brains" are in its operational language of physics; and the "symbolic language of mathematics" is another machine language" of our computational brains for the specific expressions of the underlying physics of some of the phenomena it processes. This rapidly developing language, presently much of it in "appearances" strange to the physical reality, and taken by many great minds, Plato, Galileo, Kant, Winger and likes, as indication of an incontrovertible and mysteriousness mathematical world, is in fact inadvertently, blazing the trail for discoveries of more novelties of the physics of the physical universe. This claim is based on 1) that brains only sense the world physically and operate physically and 2), that as known, on occasions, mathematics has been predictive of physics and the latter of new frontiers in the former.

Background

Mathematics, as many in the philosophical and scientific community have believed, "is the language in which the fundamental laws of physics and nature are written". Of course, mathematics in its own right has been flourishing in its many fields, generally abstract, some of which may hold the key to the further understanding of the reality of our existence; a possibility based on the past developments in areas such as group theory, topology, symmetry, and complexity mathematics, which have proved instrumental in the revelation of the physics principles. And Physics is yet to benefit from mathematics such as the Poincare conjecture [1], as well as other mathematical theorems, which the difficulty of application, communication, or unfamiliarity, may be hampering it. However the idea of the "other reality" put forward by Plato, more than two millennia ago; hinted upon in his cave allegory; and the proclamations by great minds of past centuries, the likes of Galileo and Kant, to the effect that "the essences of existence being written in the language of mathematics;" and about the "mathematics' proven effectiveness, in its applications to different fields, specifically in physics, regarded mysterious [2], have promoted the idea, by a number of scientists, that perhaps the "existence" all together, has a mathematical nature. The outspoken scientist, Max Tegmark [3] puts it this way "…our physical world not only is described by mathematics, but that is mathematics." The idea has been masterfully explored in Tegmark's book [3], deploying up-to-date scientific understanding of the matter (the Theories of Relativity, and Quantum field, and the Standard Particle Model) that only capable minds like him would allow. The beauty of the geometrical forms (old and mathematically generated) in relation to physics, and his explanation of physical phenomena as mathematical patterns, makes the idea very appealing. Certainly, as difficult matters such as the consciousness [4], and all that relate to the matters of living, become more addressable scientifically, it will become harder not to contemplate on the possibility of the beautiful idea. However, since the premise mainly draws upon the presumption of mathematics as (anthropocentrically) the language of nature, it not being the case, as will be shown, preempts the idea all together.

Another mathematical and more tangible idea of the "other reality", is the hypothesis of the existence as "virtual reality," put forward by Nick Bostrom [5]. This hypothesis *upends* the idea of the illusory world, persisting today, which has been pondered by the thinker's of the past ages; and here is a poem from the philosopher, mathematician and poet Omar Khayyam that speaks to it:

*"For in and out, above, about, below,*

*'Tis nothing but a magic Shadow-show*

*Play'd in a Box whose Candle is the Sun*

*Round which we Phantom Figures come and go"*

Obviously, the idea of everything being simulations, seemingly "a rose by any other name," is more logical, beautiful and thought provoking. And of course embedded in it, is the known fact that existence, all together, virtual or not, is brains' simulation in the final analysis; a brain capability to which dreams bear solid evidence. It is important to observe that the brain simulated worlds of the "physical beings"—regardless of possibility of the "other reality"-- must correspond to some measure of the realities of the "physical world" that they inhibit, by the dictum of the survival principle.

The concept of virtual existence is by no means far-fetched and one can envision its possibility considering the advances in computations, and particularly in the development of human-like intelligence, and the foreseeable creation of primitive Robotic consciousness; and the very realistic simulations in computer games and virtual reality applications. Therefore, if and when the technological progress in ours, or other possible civilizations, reaches certain heights -- extinction on various accounts may not allow it—the creation of a virtual reality becomes conceivable; and that its virtual occupants, with some measure of consciousness, could conduct seemingly realistic lives, the nature of which they may also question. *Given the possibility of more advanced civilization having developed such capabilities, there is a chance that our "other reality" could be a "virtual reality."* Considering that our technological progress in the future may give us the ability of creation of a virtual reality, it is conceivable also that our creator may have been a virtual reality herself. However, this makes it apparent that regardless of how many generations of virtual existence, down the line, a virtual world is, the need for a physical creator arises unless the first virtual creator happens to be Aristotle's God? While this last proposition is hard to accept, we need to accept that presence of at least one "physical reality" is a necessity from the perspective of this hypothesis.

The Thesis

While both ideas, discussed above, rely on the existence of a "physical world" for which "other reality" is hypothesized, they (seemingly) create a sense of uncertainty about the ultimate reality of our existence: virtual creation of a sophisticated technology, or inherently mathematical (forms, patterns and constructs); the possibilities that may loom in the minds. But regarding the former, the matters of biological evolution, and the cultural, social historical processes involved in the creation of the "virtual reality," all seemingly insurmountable, make the likelihood of us having physical existence much more. Therefore, with ground level certainty of physicality of our world we need first to find out how, and in what form this reality reveals itself: *it must have to do with how a truly physical being perceives and discerns her inner and outer environment*; and become aware of it as we are. The process involves the stimulation of senses, rendering brain activities (operations), which result in the recreation of the stimuli (environment simulation), and its final relay to the physiological interfaces of the body; as thought, speech and or muscular-skeletal expressions. Regarding the nature of the brain operation, it has been established by grand neurosciences research efforts to be computational [6]; and that it has been accepted at the highest levels of physics [7]. As to how brain computes, the scientific neural network computing devices [8], inspired by the brain structure, and widely deployed in artificial Intelligence (AI), have provided some clues: on the face of it, similar trial and error learning and memory development happens in both—in case of the latter it is the signal weights. Therefore, when addressing the functional operation of the brain it is not farfetched that possibly the central nervous system operates in the manner of neuronal network (parallel) computational machinery with distributed memory: A conjectured similarity , *from a ground-level computational perspective*,  that can be somewhat strengthened by drawing upon Ockham's razor, and the similitude principle [9]; and the fact that the brain and the (brain-inspired) scientific neural network both avoid the fundamental incommutability flaw inherent in the axiomatic mathematics, which digital computations in the scientific usage is afflicted with [10] The following paragraph from D. R. Hofstadter [11], speaks to the above understanding of the brain's computations, however strong in its wider connotations:

*"For another way of modeling mental processes computationally, take the neural nets—as far from the theorem proving paradigm as one could imagine. Since the cells of the brain are wired together in certain patterns, and since one can imitate any such pattern in software—that is, in a "fixed set of directives—a calculating engine's power can be harnessed to imitate microscopic brain circuitry and behavior."*

The computational operations of the brains arise from the humans' constant exposure to varying existence issues, events and complexities thereof, much sublime, occurring during the conduct of their lives. Outside of internal physiological demands, they generally relate to what their individual drives for survival engenders at the time. However, the force and the material fields, and the flood of data in the universe, are likely to be partly behind the constant activation of the central nervous system, with the brain at the helm--lifelong data churning computational machinery. Regardless of nature, complexities are sensed through the tactile sensations [12] of the five senses; detecting their streaming energy pulses, which stimulate nerves *electrochemical signaling in response to their energy characteristics*. Such signals, in totality, depending on how they make it through the hurdles of synapses to reach a neuron cell, determine the cell's measured participation in the grand complex scheme of the brain's computations, which is announced by an appropriate (electrochemical) signature down its Axon-- this signature for each neuron is the results of stochastic play out of multitudes of the afferent signals conditioned with brain's distributed electrochemical substrate of memories and existing knowledge patterns. In the traditional computers, the inputs of the program language syntax, *down to the levels of the assembler language,* stimulates pulsations of *electrical activities* (processing), on a substrate of accompanying and electronic chip hacked knowledge, which result in infinitude outputs of electric pulsations which are generally interpreted at the machine interfaces in symbolic languages-- both machinery *operations of the traditional computers and the brain are all governed by physical laws*!" In the case of the human brain computations, the outputs are resolution of complexities in numerous analogue signals that appear at body interfaces: *emanations some of which are conscious due to the availability of means of expression, such as spoken and written languages, and Arts, etc.; and some with no means, and some for which incomplete or developing mean exist.* Subjective part of consciousness and mathematics are such of the two latter cases: the former suffers from the inadequateness of expressions of the spoken language; and the latter is a developing language, and behind its progress is the gradual resolution of the ever (brain) posed complexities of the eventful phenomena of the physical universe by some well configured brains. Development of the language of mathematics, that is, the transfiguration of electrochemical signals into syntactic language in which the beauty of the formalisms of the discovered laws of physics have taken life, and its other exposes, abstract or otherwise, must have followed the manner of the development of the spoken language; and that it may be considered complementary to it, towards perfecting the natural language. Whether spoken language is internal [13] or learned, is a contentious topic among linguists; but regardless, communications necessity in time, over perhaps a few hundred millennia, must have played a significant role in shaping some of the complexity resolution electrochemical signals into its present syntactical form [4] Emphatically, mathematics is a symbolic conversion of the electrochemical output signals of specific physical complexity played in the brain; and its origin perhaps goes back to the time when humans' brains engagement with some aspects of the physical universe, rendered creations that baffled them, and some that facilitated their lives. And the blossoming of the interpretations of these events of the brains, had to wait for cultural (evolutionary) brain maturation process, until a language would develop for it; a way of formalizing extensively the evolving thoughts (brain outputs} about the some of the essences of existence, and the laws governing them. This need, in the context of anthropocentric appearing efforts, had a slow beginning until its takeoff a couple of 100 years ago, during which mathematics evolved to its present day status. Mathematics, beyond what that facilitated the formulation of the physics discoveries, a*nd the beauty they portray, in its formalism, is that of a rapidly evolving symbolic language, which in semblance to the machine language of traditional computers, is a brain computer language that represents the results of the purely physics based*

*operations of the brains.* In such a context perhaps the spoken language comparatively, can be called a higher level language. The grand attempt of grounding mathematics in logic [14], despite the near failure [10], still may speak to the similarity of the process of its development to that of the spoken language, as well as their relationship.

It is important to note that *the physics based operations of the brain* (besides it task of homeostatic upkeep), is behind all the mental efflorescence of ideas, concepts, discoveries and creativities, which are stimulated by the exposures of senses to various environmental contents and triggers; all tactile energy inputs, whether they are modulated photons, or from other methods of communication, originating from physical activity in some brains, displayed through the physiological interfaces in various layouts. Mathematics, from its very early appearances, has always had, and much more so now, findings which appeared independent of physical reality, indicative of a mysterious life of its own. As to kinds of the specific energy that triggers its evolvement, one cannot know, but based on the physics based operation of the brain, one can speculate that certain physics is behind it; and evidences of math leading physics, in the past; and new physics findings, such as supersymmetry, leading math, are perhaps important indications.

What may have led to the thoughts of mathematics being the "other reality" may have had its origin in the non physical appearances of mathematical language, symbolic in various forms, with no indication of the nature of the *under-the-hood electrical operations*, as one generally does in the case of the traditional computers, and would, even very naturally, in case of the *brain's electrochemical operations*. Of course this supposition applying more in the past, explains why the laws of mathematics have been believed to be the laws of the operations of the brain; and the following quote from Kant [15] implies such a take:

***"Did the sensations of themselves, spontaneously and naturally, fall into a cluster and order, and so became perception? No…putting sense into sensation requires innate knowledge…and because they are a priory, their laws, which are the laws of mathematics, are a priory, absolute and necessary."***

Existence is a cavalcade of innately mysterious forms with energy substrates, --Aristotle puts it as "Form is not merely shape but the shaping force"-- of which beings are a part. And brains are ever engaged in their demystification, as senses relay their stimulations by the flood of the streaming data in the environment; and the result, depending on how meaningfully they jive with the resolution potential of the brain, could be varying; from nonsense to the "Aha' moments of symbolic outcomes of mathematical discoveries, or any other symbolic representations. In essence, all well posed Problems (to discipline trained minds with intentional appearances, or by nature), in the context of brains computational ability, get some sensible results, some of which have rendered the state of human civilization; and some which will disclose more of the complex reality of the existence. And this reality is that of a "physical one," the laws of which are revealed in mathematical language formalism, though beautiful, certainly are not indicative of mathematics being its "other reality."

Conclusion:

The ideas of the "other reality," being mathematical, either as constructs or patterns, or virtual, as profound and thought provoking as they are, remain beautiful philosophical speculations that need to be taken very seriously, as any philosophical topic should. This work, while not rejecting the possibilities of virtual reality, has attempted to provide logical, and closely fact based arguments, to prove that wherever physical laws rule, the existence, perceived by physical brains, has to be truly physical. The core of the arguments is the fact that our mental mathematical outputs are just making of a language, perhaps at the level of the assembler language in the digital computers, which similarly represent the physical reality of operation of our brains' computational machinery.